\documentclass[12pt]{article}
\usepackage{graphics}
\makeatletter

\@addtoreset{equation}{section}
\makeatother

\begin{document}

\title{Simulation of I-V Hysteresis Branches in An Intrinsic 
Stack of Josephson Junctions in High $T_c$ Superconductors}

\author{ Hideki MATSUMOTO, Shoichi SAKAMOTO and Fumihiro WAJIMA \\ 
{\normalsize 
	Department of Applied Physics, Seikei University}\\  
{\normalsize 	Kichijoji Kitamachi 3-3-1}\\ 
{\normalsize    Musashino-shi, 180 JAPAN}\\ 
	and \\ 
	Tomio KOYAMA\\ 
{\normalsize Institute for Materials Research, Tohoku University}\\ 
{\normalsize Katahira 2-1-1}\\ 
{\normalsize Sendai, 980 JAPAN}\\ 
	and \\ 
	Masahiko MACHIDA \\
{\normalsize Center for Promotion of Computational 
Science and Engineering, }\\ 
{\normalsize Japan Atomic Research Institute, 2-2-54 Nakameguro,}\\ 
{\normalsize Meguro-ku, Tokyo 153, Japan}
}
\date{ }
\maketitle 
%\vskip 3truecm 
\begin{abstract}
	I-V characteristics of the high T$_c$ superconductor Bi$_2$Sr$_2$Ca$_1$C$_2$O$_8$ shows a strong hysteresis, producing many branches. 
The origin of hysteresis jumps is studied by use of the model of multi-layered Josephson junctions proposed by one of the authors (T. K.). 
The charging effect at superconducting layers produces a coupling between 
the next nearest neighbor phase-differences, which determines the structure of hysteresis branches. 
It will be shown that a solution of phase motions is understood as a combination of rotating and oscillating phase-differences, and that, at points of hysteresis jumps, there occurs a change in the number of rotating phase-differences. 
Effects of dissipation are analyzed. 
The dissipation in insulating layers works to damp the phase motion itself, while the dissipation in superconducting layers works to damp relative motions of phase-differences. 
Their effects to hysteresis jumps are discussed.
\end{abstract}
\bigskip

%	PACS number:\begin{minipage}[t]{4truecm}
%			71.27.+a\\ 
%			74.20-z\\ 
%			74.20.Mn\\ 
%			74.25.Jb\\
%			\end{minipage}\bigskip 

e-mail:matumoto@apm.seikei.ac.jp

\thispagestyle{empty} 
\newpage 
\section{Introduction}
	I-V characteristics of the high T$_c$ superconductor Bi$_2$Sr$_2$Ca$_1$C$_2$O$_8$ (BSCCO) shows a strong hysteresis, producing many branches \cite{Muller1, Muller2, Muller3}. 
Many experiments have been reported to investigate properties of hysteresis branches \cite{Sakai1,Yurgens,Tanabe1,Itoh,Helm,Sakai2,Schlenga}, and it has turned out that there is rich physics in properties of I-V characteristics. 
Substructures are found in hysteresis branches, and are discussed in relation with phonon effects \cite{Helm}. 
The d-wave effect is also discussed in higher voltage region, where the quasi-particle tunneling current becomes important \cite{Tanabe1, Schlenga}. 

The present paper is concerned to the origin of hysteresis jumps forming the main branches. 
Instead of attributing the origin of hysteresis branches to inhomogeneity of critical currents, there are several theoretical attempts to regard its origin as an intrinsic property of strong anisotropy in high T$_c$ superconductors. 
High T$_c$ superconductors have a layered crystal structure. 
Specially, BSCCO has a strong anisotropy. 
Then a model of multi-layered Josephson junctions, which has an alternating stack of superconducting and insulating layers, is considered as a proper model to describe electronic properties of those systems. 
One of main differences from the case of a single junction is the appearance of an interlayer coupling among phase-differences. Since the thickness of the superconducting layer is of the order of 3-6\AA, influence among adjacent junctions is not negligible. 

To explain the main branch structure, the following mechanisms are proposed. 
A mechanism of the interlayer coupling induced by the charging effect at the superconducting layer has been proposed by one of the author (T. K.) \cite{Koyama1}, and a branch structure of I-V characteristics is obtained \cite{Machida}.
It is pointed out in ref. \cite{Ryndyk1} that a nonequilibrium effect in superconducting layers also mediates an inter-layer coupling, and I-V characteristics are discussed \cite{Ryndyk2, Sakai2}. 
When the magnetic field is applied, the inductive coupling becomes important, and its effects have been discussed \cite{Sakai1,Kleiner,Bulaevskii, Tachiki, Pederson}. 

In this paper, we investigate in detail the scheme of hysteresis jumps appearing in main branches without magnetic filed. 
Since the inductive coupling among adjacent junctions does not exist, the charging effect at superconducting layers becomes important. 
A charging at a superconducting layer induces the change of electric fields in the neighboring insulating layers and induces an inter-layer coupling, since the time-derivative of the phase-difference is related with the voltage. 
We will show that the origin of the multi-branch structure is the nonlinear effect among phase-differences through the charging effect. 
Various disturbances, such as dissipation in insulating layers, dissipation in superconducting layers, noise, a surface proximity effect affect schemes of hysteresis jumps. 
We will investigate how the schemes of hysteresis jumps are affected by the above mentioned disturbances. 
The dissipation in the superconducting layer will be introduced phenomenologically through the charge relaxation. 
Schemes of hysteresis jumps are obtained by increasing applied current gradually up to certain value and then decreasing. 
Obtained equations show that the resistive dissipation in the insulating layer works to damp the motion of phase-difference in each junction, while the charge relaxation in the superconducting layer works to damp relative motions of phase differences among neighboring junctions. 
This fact affects the way of formation of hysteresis jumps in the process of increasing and decreasing applied current. 
It will turn out that the dissipation in superconducting layer and noise enhance the occurrence of jumps in both increasing and decreasing current, although the main branch structure is determined by the coupling through the charging effect. By comparing the result of numerical simulation with the experiment of a small stack \cite{Itoh}, we will show validity of the present model. 

In the next section, the necessary formula for calculation is summarized. 
The dissipation in superconducting layers is introduced phenomenologically. 
In Sect. 3, results of numerical simulation are presented. 
The origin of hysteresis jumps is investigated. 
It will be shown that, at a position of a jump, a stationary solution changes its form. 
We present a comparison with the experiment of ref. \cite{Itoh} and show validity of the charging mechanism. 
Sect. 4 is devoted the conclusion. 

\section{Josephson Junction Stacks}
	In this section we summarize the multi-Josephson junction model of high T$_c$ superconductors \cite{Koyama1}. 
Let us consider the $N+1$ superconducting layers, numbered from $0$ to $N$. We denote the gauge invariant phase difference of the $(l-1)$-th and $l$-th superconducting layer by $\varphi(l)$, and voltage by $V(l)$. 
The widths of the insulating and superconducting layers are denoted by $D$ and $s$, respectively. 
At the edges, the effective width of the superconducting layer may be extended due to the proximity effect into attached lead metals. 
The widths of the $0$-th and $N$-th superconducting layers are denoted by $s_0$ and $s_N$, respectively. 

	We assume that physical quantities are spatially homogeneous on each 
layer. This assumption is applicable in cases of no applied magnetic field and 
small sample size in the a-b direction.
According to the procedure presented in ref. \cite{Koyama1}, we have the following equation for the total current $J$ flowing through each junction as 
\begin{equation}
	\frac{J}{J_c}=j_c(l)\sin\varphi(l)+\beta \frac{V(l)}{V_0}
		+\frac{1}{\omega_p}\frac{\partial}{\partial t}\frac{V(l)}{V_0} \ .\label{J} 
\end{equation} 
The current $J$ is normalized by the critical current $J_c$, $j_c(l)=J_c(l)/J_c$  with $J_c(l)$ being the critical current for the $l$-th junction. The time $t$ is normalized by the inverse of the plasma frequency, $\omega _p$ = $\sqrt{\frac{2e}{\hbar}\frac{4\pi DJ_c}{\epsilon}}$, with $\epsilon$ being dielectric constant of the insulating layer. The voltage $V(l)$ is normalized by $V_0=\frac{\hbar\omega_p}{2e}$. 
The parameter $\beta$ is given by  $\beta=\frac{\sigma V_0}{J_cD}$. 
In this paper we assume
\begin{equation}
	j_c(l)=1\ \mbox{\rm for\ all\ }l\ ,
\end{equation}
for simplicity. The relation between time-derivative of phases and voltages at junctions is given by 
\begin{equation}
	\frac{1}{\omega_p}\frac{\partial}{\partial t}\varphi(l)
	=\sum_{l'=1}^NA_{ll'}\frac{V(l')}{V_0} \ . \label{dtvarphi}
\end{equation}
The matrix $A$ is given by
\begin{equation}
	A=\pmatrix{ 
	1+\alpha(1+\frac{s}{s_0}) &-\alpha 	&0 		&\cdots	& 
								& \cr   
	-\alpha	                  &1+2\alpha	&-\alpha	&0    	&\cdots
								& \cr 
	0                         &-\alpha	&1+2\alpha	&-\alpha&0
								&\cdots \cr   
				  &		&\cdots		&	&
								&\cr   
	\cdots			  &0		&		&	&-\alpha					&1+\alpha(1+\frac{s}{s_N})\cr  }\ ,
       \label{A}
\end{equation}
where  
\begin{equation}
	\alpha=\frac{\mu^2\epsilon}{sD}      \label{alpha}
\end{equation}
with $\mu^2$ being given by the relation between the charge density and scalar potential $A_0$ as \cite{Matsumoto} 
\begin{equation}
	4\pi\rho=-\mu^{-2}(A_0
		+\frac{\hbar }{2e}\frac{\partial}{\partial t}\phi) \ ,
\end{equation}
where $\phi$ is the phase of the order parameter defined on the superconducting layer.
In the present analysis, we assume that the current flows only along the c-axis, say, the z-direction. 
Since the current $j_z$ in superconducting layers is given by
\begin{equation}
	\frac{4\pi}{c}j_z
	=-\lambda^{-2}(A_z-\frac{\hbar c}{2e}
                       \frac{\partial}{\partial z} \phi) \ ,
\end{equation}
the current conservation law leads to
\begin{equation}
	\mu^{-2}\frac{1}{c}\frac{\partial}{\partial t}A_0
	+\lambda^{-2}\frac{\partial}{\partial z}A_z=0\ , \label{gc}
\end{equation}
under the condition
\begin{equation}
	\left(\mu^{-2}(\frac{1}{c}\frac{\partial}{\partial t})^2
	-\lambda^{-2} \Bigl(\frac{\partial}{\partial z} \Bigr)^2
               \right)\phi=0\ .
\end{equation}
The gauge condition (\ref{gc}) is called phason gauge \cite{Matsumoto}, and is determined in such a way that the response of the current and charge to the space-time variation of phase appears in a gauge invariant form.
It is shown in ref. \cite{Ryndyk1} that a nonequilibrium effect caused by the tunneling current induces the dissipation effect in $\mu^{-2}$. 
Since the superconducting layer is thin, the current also receives effects from boundaries between superconducting and insulating layers, which may also works as dissipation to the superconducting current. 
When the effect of the d-wave superconductivity is considered, one expects also certain dissipation due to the presence of the gap-less region. Such effects have not been much studied yet. Then we introduce phenomenologically the dissipation in superconducting layer in a Lorentz form as 
\begin{equation}
	\mu^{-2}=\mu^{-2}_0\frac{1}{1-i\omega\tau}\ .
\end{equation}
This corresponds to the following phenomenological form of charge relaxation, 
\begin{equation}
	\frac{\partial}{\partial t}\rho
	=-\frac{1}{\tau}(\rho+\frac{\mu_0^{-2}}{4\pi}(A_0+
            \frac{\hbar }{2e}\frac{\partial}{\partial t}\varphi))\ .
\end{equation}
By considering the expression $\alpha$ in Eq. (\ref{alpha}), we replace $\alpha$ by 
\begin{equation}
	\alpha\rightarrow\alpha(1-i\omega\tau)
\end{equation}
with $\alpha$ being given by Eq. (\ref{alpha}) replaced $\mu^2$ by $\mu^{2}_0$. 
From Eqs. (\ref{J}) and (\ref{dtvarphi}), we have 
\begin{equation}
	\frac{1}{\omega_p^2}\frac{\partial^2}{\partial t^2}\varphi(l)
	=\sum_{l'}(A_{ll'}+A_{1ll'}\tau\frac{\partial}{\partial t})
	(\frac{J}{J_c}-j_c(l')\sin\varphi(l'))
	-\beta\frac{1}{\omega_p}\frac{\partial}{\partial t}\varphi(l), 
	\label{dtvarphi1}
\end{equation}
where $A$ is the matrix $A$ in Eq. ($\ref{A}$) and $A_1$ is given as 
\begin{equation}
	A_1=\pmatrix{ 
	\alpha(1+\frac{s}{s_0}) &-\alpha 	&0 		&\cdots	& 
								& \cr   
	-\alpha                  &2\alpha	&-\alpha	&0    	&\cdots
								& \cr 
	0                         &-\alpha	&2\alpha	&-\alpha&0
								&\cdots \cr   
				  &		&\cdots		&	&
								&\cr   
	\cdots			  &0		&		&&-\alpha					&\alpha(1+\frac{s}{s_N})\cr  }\ .
       \label{A1}
\end{equation}

As the above derivation shows, the charging effect at superconducting layers is the main mechanism inducing the inter-layer coupling through the electromagnetic interaction, when the spatial homogeneity in the a-b plane is satisfied. 
A charge at a superconducting layer modifies electric fields in the neighboring insulating layers, and the change of electric fields affects the time-derivative of phase differences. 
Phonon effect considered in ref. \cite{Helm} may be introduced through the frequency dependence of the dielectric constant $\epsilon$, which is not considered in this paper.
The equation (\ref{dtvarphi1}) means that the dissipation in the insulating layer works as a damping of phase motions at each junction, and that the dissipation in the superconducting layer works as a damping of the relative phase motion. 
In the next section, we present results obtained by numerical simulation of Eq. (\ref{dtvarphi1}). 

\bigskip

\section{Results of Numerical Simulation}
We have performed numerical simulations of Eq. (\ref{dtvarphi1}) by use of the 5-th order of the predictor corrector method. 
In all calculation, we choose the parameter $\alpha=1.0$, $\beta=0.2$. The time step $dt$ is chosen as $\omega_pdt=1.0\times 10^{-3}$. \par 
The parameter $\beta$ is related to the conductivity of the insulating layer. 
Although it works to lead a system into a stationary state and determines the slope of the I-V characteristics, its effect to a scheme of hysteresis jumps is not sensitive in this parameter range. 
The parameter $\alpha$ plays an important role to determine hysteresis branches due to non-linear effects among phase-differences. 
The value of $\alpha$ in BSCCO was estimated in ref. \cite{Koyama1} as $1<\alpha<3$ and, by use of $\epsilon=25$ for La$_{2-x}$Sr$_x$CuO$_4$, the value of $\alpha=2.7$ was obtained. 
There is a discussion that $\epsilon$ for BSCCO may be smaller, and from a microscopic theory, $\alpha$ is estimated much smaller value as 0.2 \cite{Preis}. 
Also it was reported that the equal spacing of I-V branches observed in experiments of BSCCO is reproducible, with a smaller choice of $\alpha \leq 1$ \cite{Koyama2}. 
Here we use a typical value $\alpha=1.0$. 

%Correction 
By use of Eqs. (\ref{dtvarphi}) and (\ref{J}), the relation for obtaining a voltage for each junction is written as
\begin{equation}
	\frac{1}{\omega_p}\frac{\partial}{\partial t}\varphi(l)
	=\sum_{l'}((A-\tau\beta A_1)_{ll'}\frac{V(l)}{V_0}
		 +\tau A_{1ll'}(\frac{J}{J_c}-j_c(l')\sin\varphi(l')) )\ .
\end{equation}
The average of voltage $\bar V(l)$ is given by
\begin{equation}
	\bar V(l)=\frac{1}{T_{max}-T_{min}}\int_{T_{min}}^{T_{max}}dtV(l)\ .
\end{equation}
The total DC-voltage $V$ is obtained by
\begin{equation}
	V=\sum_{l=1}^N\bar V(l)\ .
\end{equation}

	In Fig. 1, I-V characteristics is presented for $N=10$, $s_0/s=s_N/s=2$, $\tau\omega_p=0.1$. 
The current $J$ is gradually increased with the step $dJ/J_c=0.01$ up to $J/J_c=2.0$, and then gradually decreased. 
The I-V curve shows the jump at $J/J_c=1.0$ and, after two jumps, it linearly increases up to $J/J_c=2.0$. 
In the current decreasing process, there appear three jumps when the current becomes smaller than $J/J_c=0.4$. 

	In Fig. 2, we show the branch structure in the I-V characteristics corresponding to Fig. 1. 
This is obtained by decreasing the current when a jump occurs in the current increasing process, and by increasing the current when a jump occurs in the current decreasing process. 
It should be noted that the positions of jumps labeled by 1 and 5 lie on the same branches, while the positions 2 and 4 are close but lie on the different branches. 

	In order to understand this branch structure, we show, in Fig. 3, the voltage distributions on each junction just after the jumps. 
Fig. 3(a) is for the current increasing process and Fig. 3(b) is for the current decreasing process. 
For the line labeled 1, two junctions at the edges have higher voltage. 
The solution shows that the phases with higher voltages increases approximately linearly in time, corresponding to the phase rotating motion, while other junctions have oscillating phase motions. 
In fact, when the time average of $\frac{\partial\varphi(l)}{\partial t}$ is plotted in Fig. 4, junctions with rotating and oscillating phases are clearly identified. 
As can be seen in Fig.3, the number of junctions with rotating phase increases as 2, 4, and 6 due to the symmetric arrangement. 
The number of phase rotating junctions is two for the voltage lines labeled by 1 and 5, four for 2 and 4. 
The lines 1 and 5 shows the same pattern of voltage distribution and therefore form the same branch in Fig.3, while the line 2 and 4 show the different pattern, though the number of the phase rotating junction is same. 	

With this analysis, we see that the hysteresis jumps are associated with the change of the distribution pattern of rotating phase motions. 
If the $l$-th junction has a rotating phase, the time average of 
$\frac{\partial\varphi(l)}{\partial t}$ is constant and that of $\sin(\varphi(l))$ is zero. 
If the $l$-th junction has an oscillating phase, the time average of $\frac{\partial\varphi(l)}{\partial t}$ is zero and that of $\sin(\varphi(l))$ is constant. 
Let us consider that $N_R(\geq 0)$ junctions have rotating phases.  
Considering the above facts, we can obtain the equation to determine the voltage-current relation from the time-average of Eqs. (\ref{J}) and (\ref{dtvarphi1}) as
\begin{equation}
	\beta \frac{V}{V_0}=N\frac{J}{J_c}
	-\sum_{l_0}j_c(l_0)<\sin\varphi(l_0)>\ ,\label{bV}
\end{equation}
\begin{equation}
	\sum_{l_0'}A_{l_0l_0'}j_c(l')<\sin\varphi(l_0')>=a(l_0)\frac{J}{J_c}\ ,
		\label{sin}
\end{equation}
where $<\cdots>$ indicates time average, $l_0$ and $l_0'$ are the labels of junctions with oscillating phases, 
\begin{equation}
	a(1)=1+\alpha\frac{s}{s_0}\ ,a(N)=1+\alpha\frac{s}{s_N} ,\  
	\mbox{\rm otherwise}\ a(l)=1\ .
\end{equation}
As is shown in Fig. 3, even if the number of phase-rotating junctions is same, slightly different branches are formed according to the pattern of distribution of rotating phases. That is, branches are grouped by the  number of phase-rotating junctions, $N_R$, and different patterns of distribution of phase-rotating junctions lead slightly shifted branches. 
The Eqs. (\ref{bV}) and (\ref{sin}) are solved in a form
\begin{equation}
	\beta \frac{V}{V_0}=n\frac{J}{J_c}\  \label{n}.
\end{equation}
In Fig. 5, we present $n$ versus $N_R$, the number of rotating phases. 
The small dots are for all possible patterns. 
The edge junctions mostly rotate, and rotating phases have mostly oscillating phases in the next neighbors. 
Taking into account these facts, we have the larger circles with the restrictions, 1) pattern is symmetric, 2) two edges have rotating phases and 3) there are no patterns with more than three successive rotating phases. 
From the figure we can see that branches appears roughly with equal spacing but that those at higher voltage distribute denser. 
In the numerical simulation, it seems that only special patterns are realized. 
The simulated result in Fig. 2 gives $n=3.21$ for $N_R=2$, $n=6.21$ and $6.47$ for $N_R=4$, and $n=8.59$ for $N_R=6$.  
What stability conditions should be imposed in addition to Eqs. (\ref{bV}) and (\ref{sin}) is still an open question. 

	In order to see how the scheme of hysteresis jump changes with various conditions, we show, in the following, structures of branches in I-V characteristics. 
In these analyses, the current is increased up to $J/J_c=2$ and decreased to zero. 
Branches are picked up when a voltage jump occurs. Fig. 6 is for finite junction stack $N=10$, and $s_0=s_N=2.0$ is chosen. 
The parameters are $\alpha=1.0$, $\beta=0.2$. 
In Fig. 6, (a) is for $\omega_p\tau=0.0$ without noise, (b) is for $\omega_p\tau=0.1$ without noise, (c) is for $\omega_p\tau=0.0$ with noise, and (d) is for $\omega_p\tau=0.1$ with noise. 
The noise is introduced through the small randomness of the time derivative of the phase at each time step, $\delta (\frac{1}{\omega_p}\frac{\partial\varphi(l)}{\partial t})$ = 
$1.0\times 10^{-4}$. 
Fig. 7 is for N=10 with the periodic boundary condition. 
Fig. 7(a) is for $\omega_p\tau=0.0$ without noise, (b) is for $\omega_p\tau=0.1$ without noise, (c) is for $\omega_p\tau=0.0$ with noise, and (d) is for $\omega_p\tau=0.1$ with noise. 

When a noise is introduced, possible patterns are not necessarily symmetric, so that the number of branches increased. 
In addition, an instability to cause a hysteresis jump is much easily enhanced. 
The effect of $\tau$ also enhance hysteresis jumps, which may be understood from the fact that it works as damping of the relative motion. 
Specially, it enhances hysteresis jumps in the current increasing process. 

In the periodic cases, this tendency is more enhanced as can be seen from comparison with Fig. 7(a) and others. 
In a long stack, where surface effects are less, we can say that the dissipation effect in the superconducting layer or noise play an important role to induce hysteresis jumps. 

In Fig. 8, we present the comparison with our numerical result and the experiment of ref. \cite{Itoh}. 
We choose $N=8$ from their arrangement of the sample. 
In order to explain large extension of hysteresis branches in the current increasing process, we assume that the critical currents $J_c$ are reduced at the edges. 
We choose $J_c=0.47meV$, since the numerical analysis shows that the $J_c$ is situated roughly at the middle of the second branch. 
From the data, we have $J_c(0)/J_c=0.4$, $J_c(N)/J_c=0.9$ and other $J_c(l)/J_c=1.0$.  
The parameter $\alpha$ is chosen 1.0, by assuming $s=6$\AA, $D=6$\AA, $\mu=$2\AA and $\epsilon=10$. 
In ref. \cite{Kadowaki}, it is reported that the longitudinal plasma energy is of order of $0.5meV$. 
Then $V_0$ is $0.25meV$. 
The $\beta$ is estimated from the branch 4; that is, the branch 4 has four rotating phases and is on the line passing $J=0.9mA$, $V=60meV$. 
In Eq. (\ref{n}) we choose $n=6$ from the result of Fig. 5, and $\beta$ is estimated as $\beta=0.05$. 
Since we do not introduce any other parameters, we set $s_0/s=1.0$, $s_N/s=1.0$. The parameter $\omega_p\tau$ is chosen as $0.5$, which is obtained by trial and error to produce as many branches as possible. 
The inset is from ref. \cite{Itoh}. 
The labels of branches correspond to those in the inset. 
The numbers in the brackets indicate the number of rotating phases.  
The values of J, at which hysteresis jumps occur, fit very well with experimental values of the inset. 
Hysteresis jumps are induced when certain instability develops in the non-linear equation of (\ref{dtvarphi1}). 
Though each junction, except for one at edges, has the same $J_c$, a critical 
$J$ for each branch is different and is controlled mainly by $\alpha$. 
The agreement of this critical $J$-values observed in Fig. 8 supports the validity of the present charging mechanism. 
In Fig. 8 certain branches in the increasing current are missing. How hysteresis jumps occurs depends of effects of disturbances. 
There may be certain noises to make development of instability easier. 
The length of branches depends on choice of parameter. 
In decreasing current, there are a few discrepancies. 
In numerical calculation, the I-V curve is straight, while the experiment shows a curvature in the low current region. 
In addition, the positions of jumps are different. 
Such discrepancy may relate effects of d-wave superconductivity, such as 
the presence of the quasi-particle tunneling, for example, but it is beyond the discussion of the present work. 

\section{Conclusion}
In this paper we have studied the I-V characteristic branches in high T$_c$ superconductors by the multi-layered Josephson junction model. 
We have shown that the charging effect in superconducting layers causes an inter-layer coupling among phase-differences and that hysteresis jumps are induced as a result of non-linear effect among phase-differences. 
The appearance of hysteresis branches is intrinsic in high T$_c$ superconductors. On this aspect, we note the experimental report that, even in a sample with negligible $J_c$-inhomogeneity, a branch structure is observed \cite{Schlenga}.

Due to the weak interlayer coupling arising from the charging effect, equations for phase-differences form a non-linear coupled differential equation. 
Its solution is classified by the number of rotating phases. 
The change of its number occurs intrinsically in this nonlinear equation, which leads to formation of I-V characteristic branches. 
Although the number of rotating phases is same, different patterns of distribution of rotating phases lead slightly shifted branches. 
It is still an open question and a future problem which patterns are more easily realized among many possible patterns. 

In this system, the resistive dissipation in the insulating layer works as damping of phase motion in each junction and the dissipation caused by charge relaxation in superconducting layers works as damping of relative phase motions. 
The effects of dissipation, noise, and surface proximity to structure of hysteresis jumps in I-V characteristics are investigated. 
The dissipation in superconducting layer and noise enhance the occurrence of jumps in both increasing and decreasing current. 

We have presented the comparison with the experiment in Fig. 8.  
The agreement of the critical $J$ for each hysteresis branch shows validity of the charging mechanism as the origin of the inter-layer coupling. 

Finally, we comment on effects of the d-wave superconductivity. 
The fact that the Josephson current is given by the phase-difference is not modified in the d-wave superconductivity with a single component. 
Therefore, the main conclusions in the present paper are not modified. 
As was pointed out in the text, the d-wave superconductivity gives, due to the presence of a gapless region, an additional contribution to dissipation in superconducting layers and additional quasi-particle tunneling current even in the low voltage region. 
Effects of such contributions may be interesting future problems. 

\bigskip
\newpage 
\begin{center}
	{\Large\bf Acknowledgements}
\end{center}
	This work was supported by Special Research Grant in Faculty of Engineering, Seikei University. 
One of the authors (F. W.) was supported by Ikuei scholarship in Japan. 

\newpage 

\noindent 
{\Large \bf Figure captions}\bigskip 

\noindent  
Fig.1
\begin{minipage}[t]{13truecm}
I-V characteristics \\ 
Parameters are $\alpha=1.0$, $\beta=0.2$, $N=10$, $\frac{s_0}{s}
=\frac{s_N}{s}=2.0$ and 
$\omega_p\tau=0.1$. 
\end{minipage}\\ \\ 
Fig.2
\begin{minipage}[t]{14.5truecm}
I-V characteristics and branch structure \\ 
Parameters are $\alpha=1.0$, $\beta=0.2$, $N=10$, $\frac{s_0}{s}
=\frac{s_N}{s}=2.0$ and 
$\omega_p\tau=0.1$. 
\end{minipage}\\ \\ 
Fig.3
\begin{minipage}[t]{14.5truecm}
Voltage distribution \\ 
Parameters are $\alpha=1.0$, $\beta=0.2$, $N=10$, $\frac{s_0}{s}
=\frac{s_N}{s}=2.0$ and $\omega_p\tau=0.1$.
(a) is for increasing current, and (b) is for decreasing current. 
\end{minipage}\\ \\ 
Fig.4
\begin{minipage}[t]{14.5truecm}
Average $\frac{1}{\omega_p}\frac{\partial\varphi}{\partial t}$ distribution \\ 
Parameters are $\alpha=1.0$, $\beta=0.2$, $N=10$, $\frac{s_0}{s}
=\frac{s_N}{s}=2.0$ and 
$\omega_p\tau=0.1$. 
(a) is for increasing current, and (b) is for decreasing current. 
\end{minipage}\\ \\ 
Fig.5
\begin{minipage}[t]{14.5truecm}
Slope $n$ vs. the number of rotating phases $N_R$\\ 
Slope $n$ is defined by $\beta\frac{V}{V_0}=n\frac{J}{J_c}$. 
Parameters are $\alpha=1.0$, $N=10$, $\frac{s_0}{s}=\frac{s_N}{s}=2.0$. 
Small dots correspond to all possible patterns, while larger circles are with restrictions presented in the text. 
\end{minipage}\\ \\ 
Fig.6
\begin{minipage}[t]{14.5truecm}
I-V characteristics and branch structure for a finite stack \\ 
Parameters are $\alpha=1.0$, $\beta=0.2$, $N=10$, $\frac{s_0}{s}
=\frac{s_N}{s}=2.0$. (a)$\omega_p\tau=0.0$ without noise, 
(b)$\omega_p\tau=0.1$ without noise, 
(c)$\omega_p\tau=0.0$ with noise, and (d)$\omega_p\tau=0.1$ with noise.
\end{minipage}\\ \\ 
Fig.7
\begin{minipage}[t]{14.5truecm}
I-V characteristics and branch structure for a periodic stack \\ 
Parameters are $\alpha=1.0$, $\beta=0.2$, $N=10$, $\frac{s_0}{s}
=\frac{s_N}{s}=2.0$. (a)$\omega_p\tau=0.0$ without noise, 
(b)$\omega_p\tau=0.1$ without noise, 
(c)$\omega_p\tau=0.0$ with noise, and (d)$\omega_p\tau=0.1$ with noise.
\end{minipage}\\ \\ 
Fig.8
\begin{minipage}[t]{14.5truecm}
I-V characteristics\\ 
Inset is from ref. \cite{Itoh}. The labels of the branches are identified 
with those of the inset. The numbers in the bracket are the numbers of 
rotating-phases.  
\end{minipage}\\ 

\begin{figure}
 \includegraphics[16truecm, 22truecm]{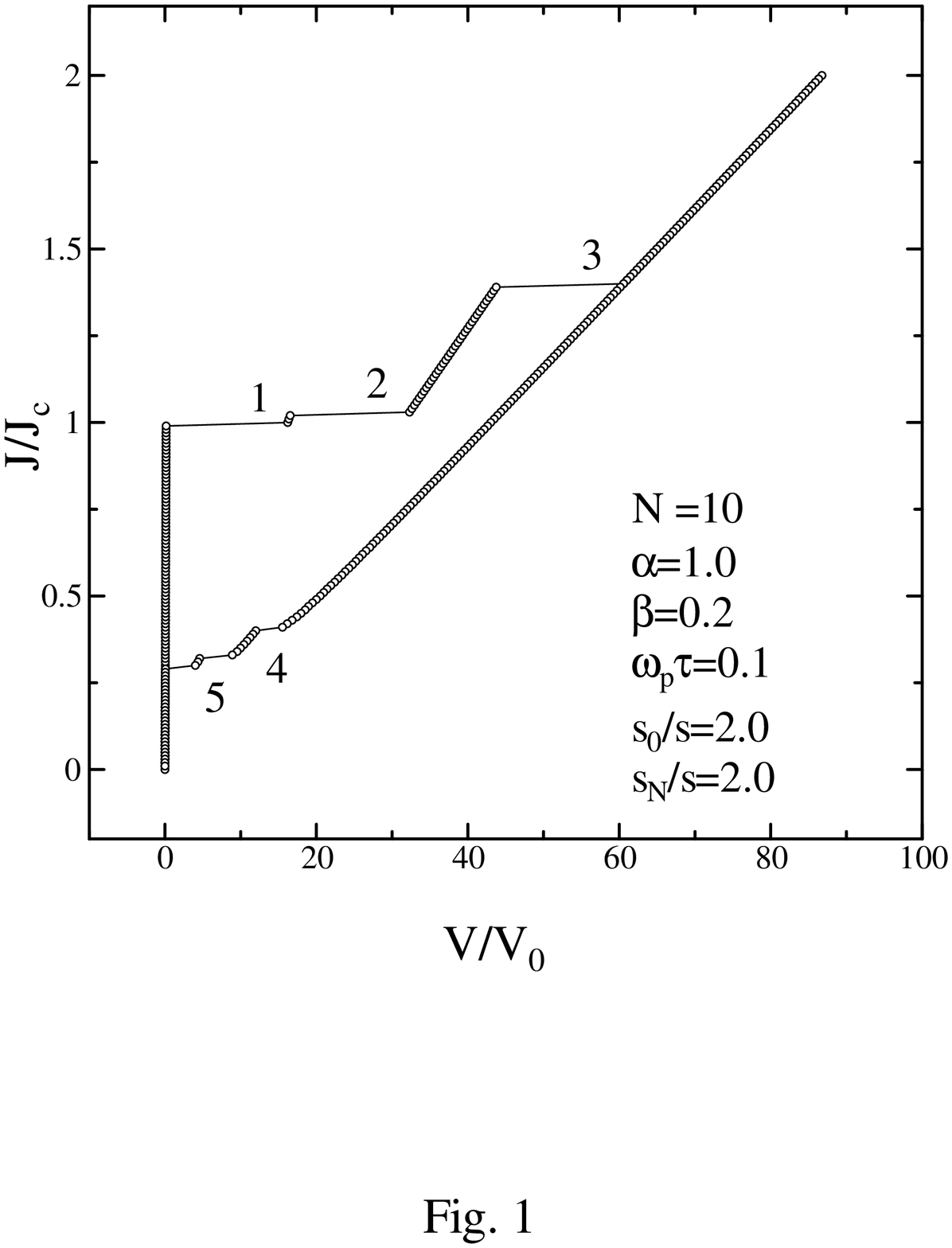}
\end{figure}
\newpage 
\begin{figure}
 \includegraphics[16truecm, 22truecm]{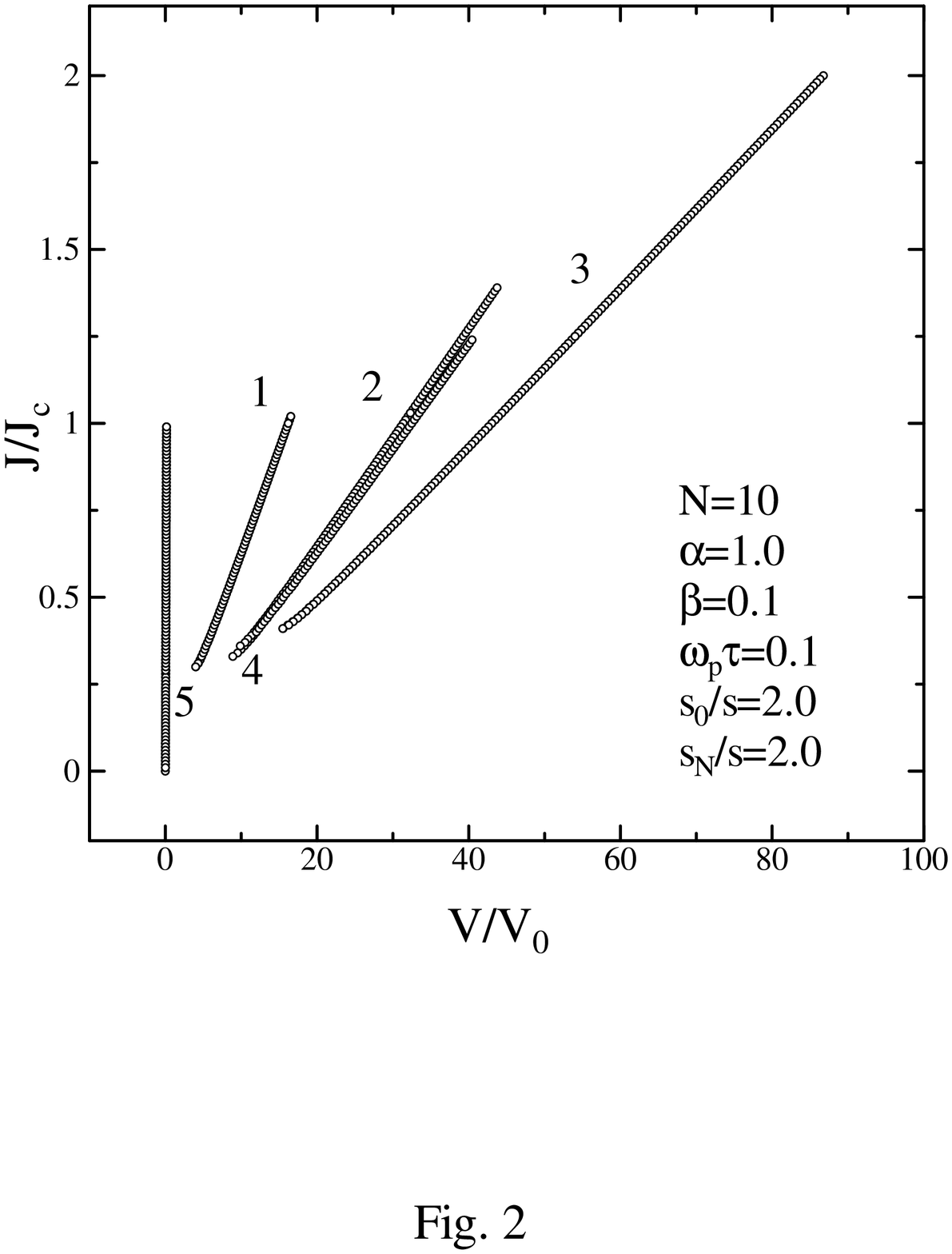}
\end{figure}
\newpage 
\begin{figure}
 \includegraphics[16truecm, 22truecm]{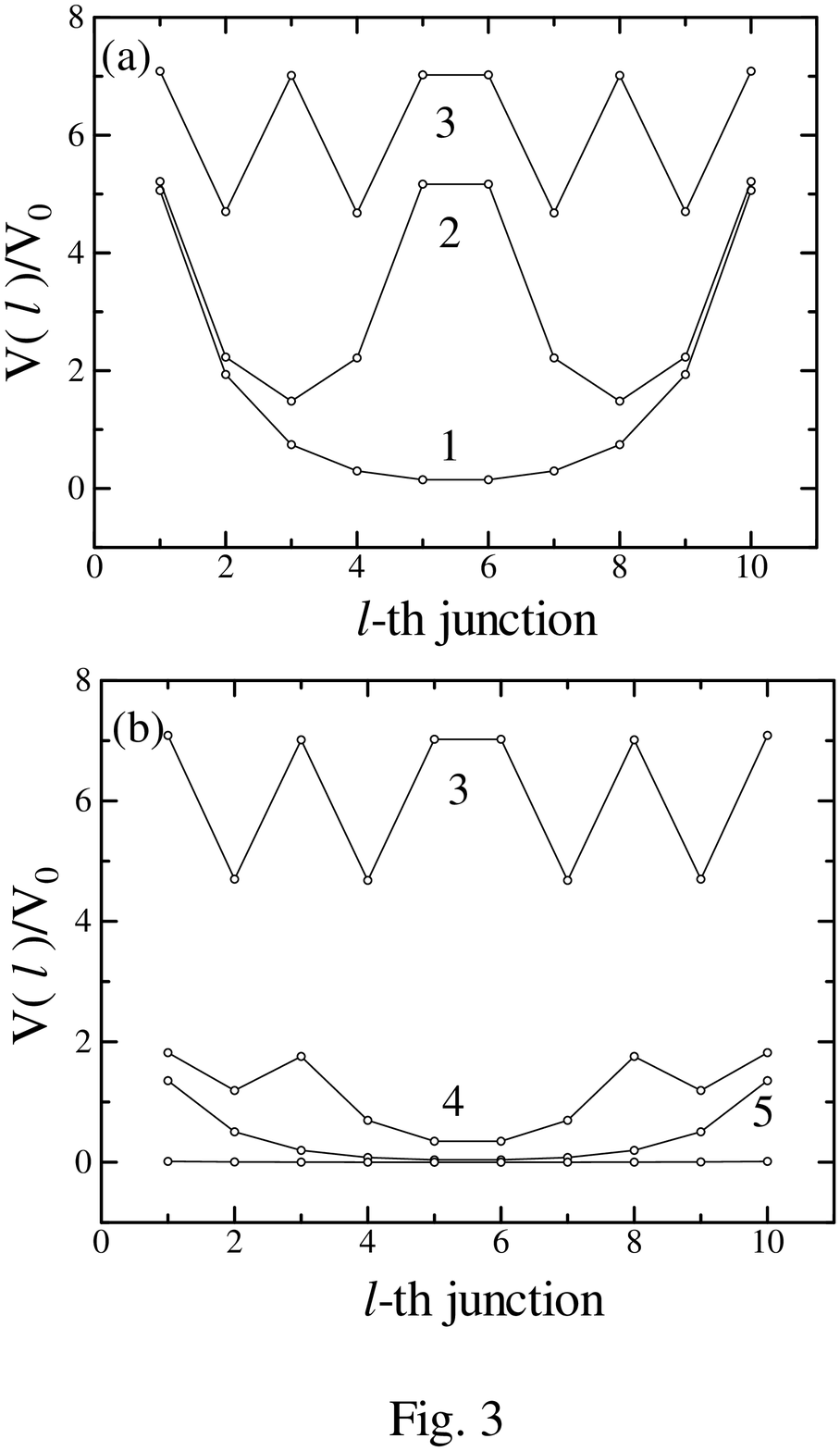}
\end{figure}
\newpage 
\begin{figure}
 \includegraphics[16truecm, 22truecm]{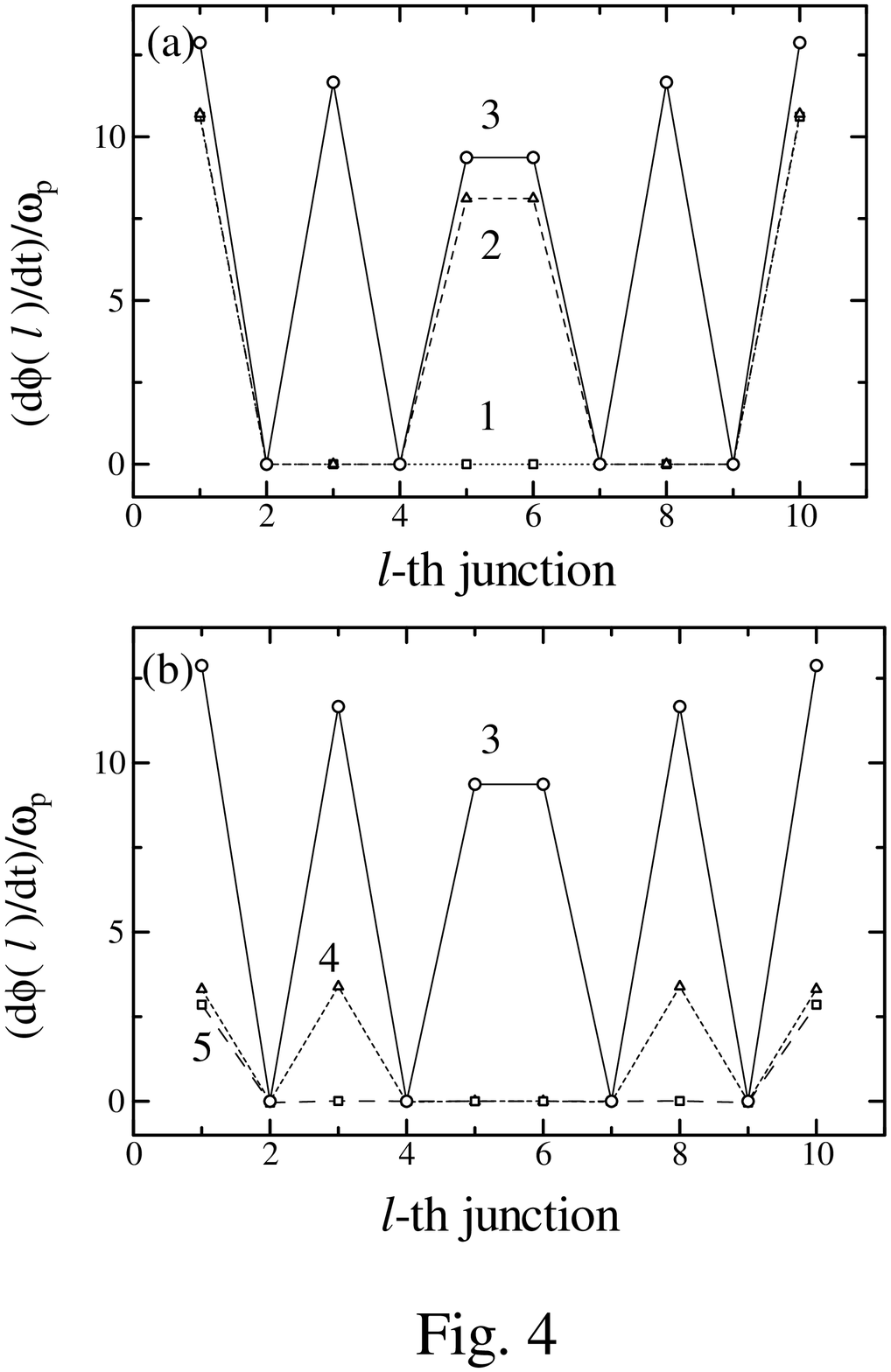}
\end{figure}
\newpage 
\begin{figure}
 \includegraphics[16truecm, 22truecm]{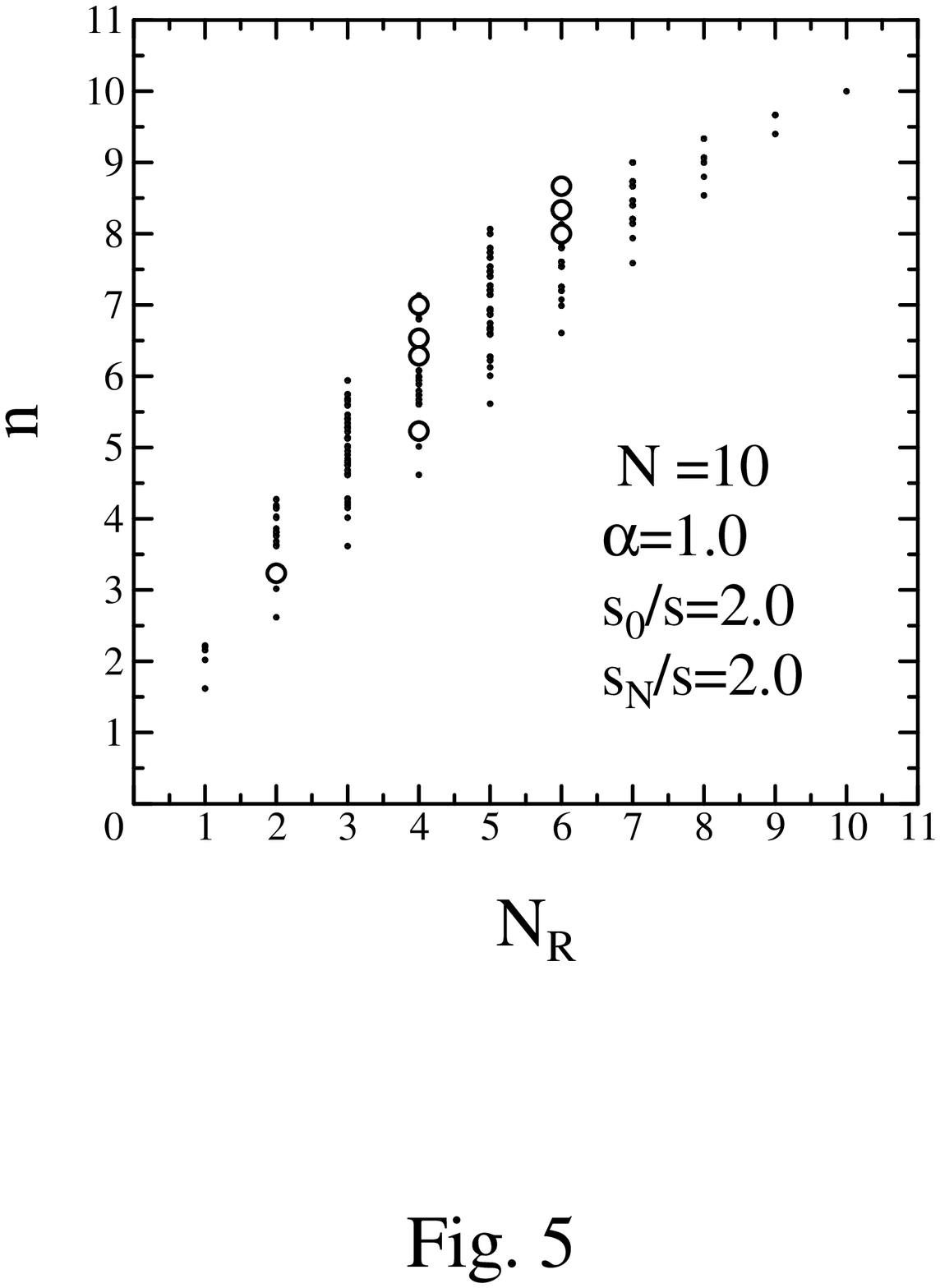}
\end{figure}
\newpage 
\begin{figure}
 \includegraphics[16truecm, 22truecm]{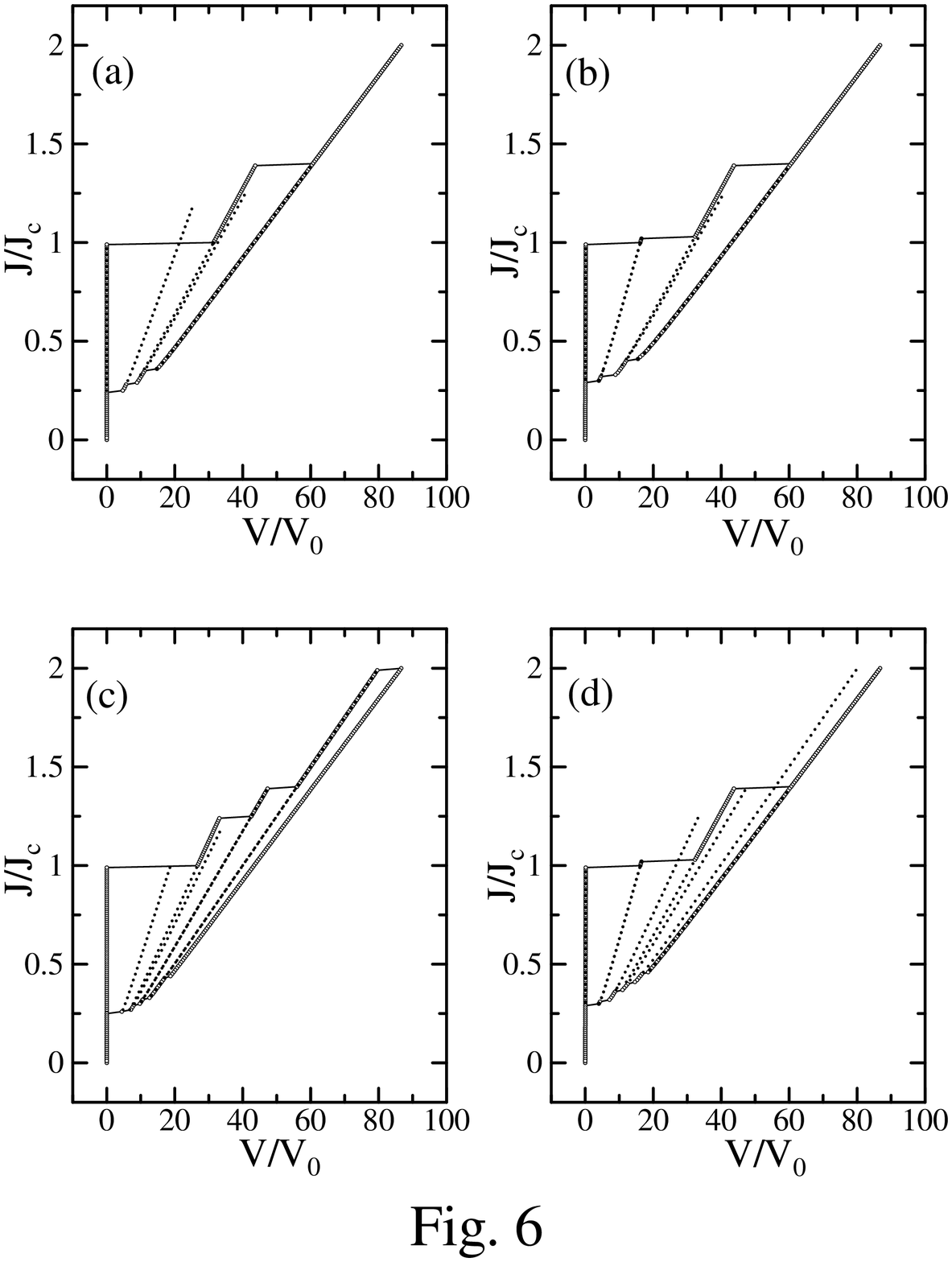}
\end{figure}
\newpage 
\begin{figure}
 \includegraphics[16truecm, 22truecm]{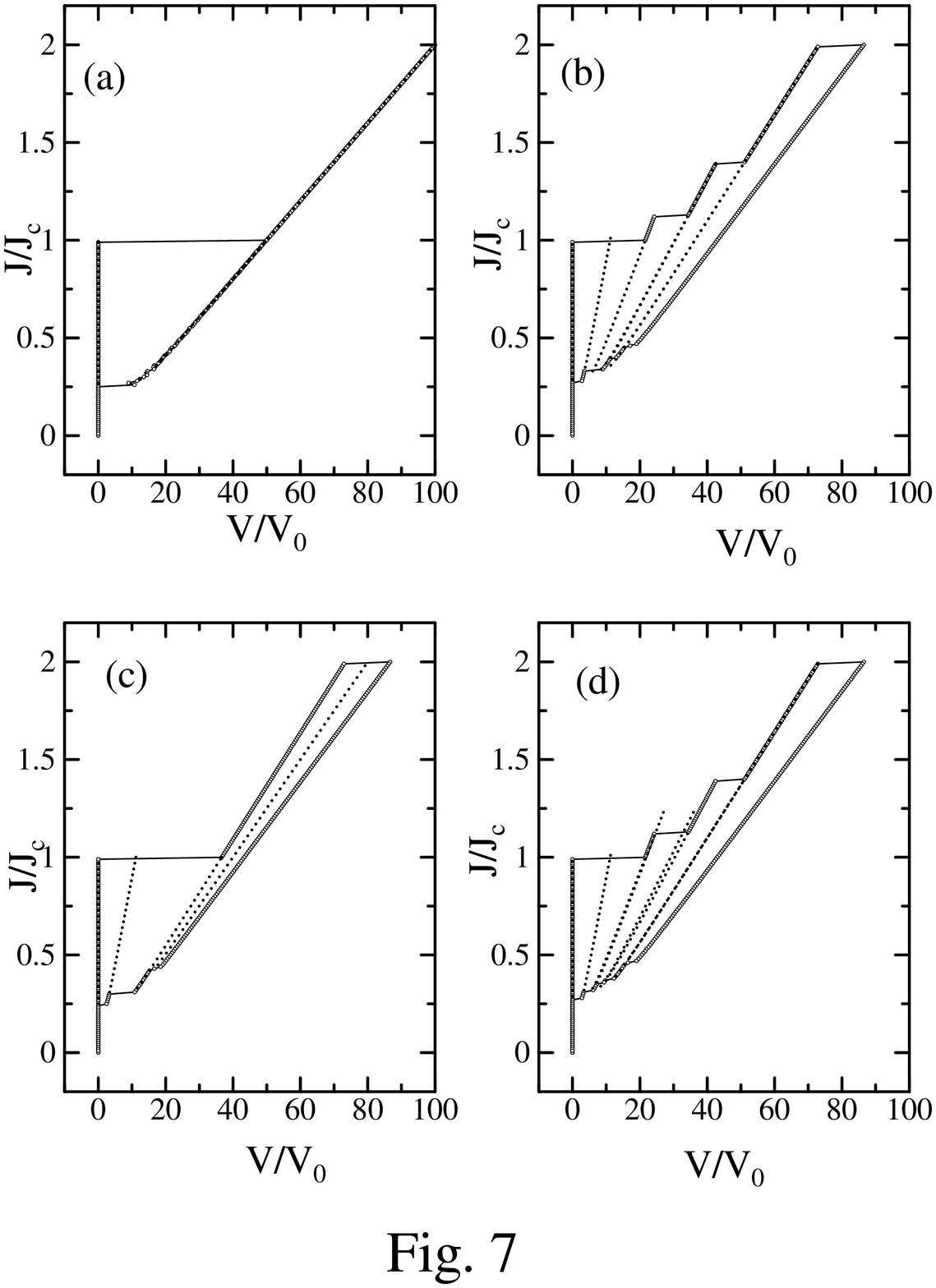}
\end{figure}
\newpage 
\begin{figure}
 \includegraphics[16truecm, 22truecm]{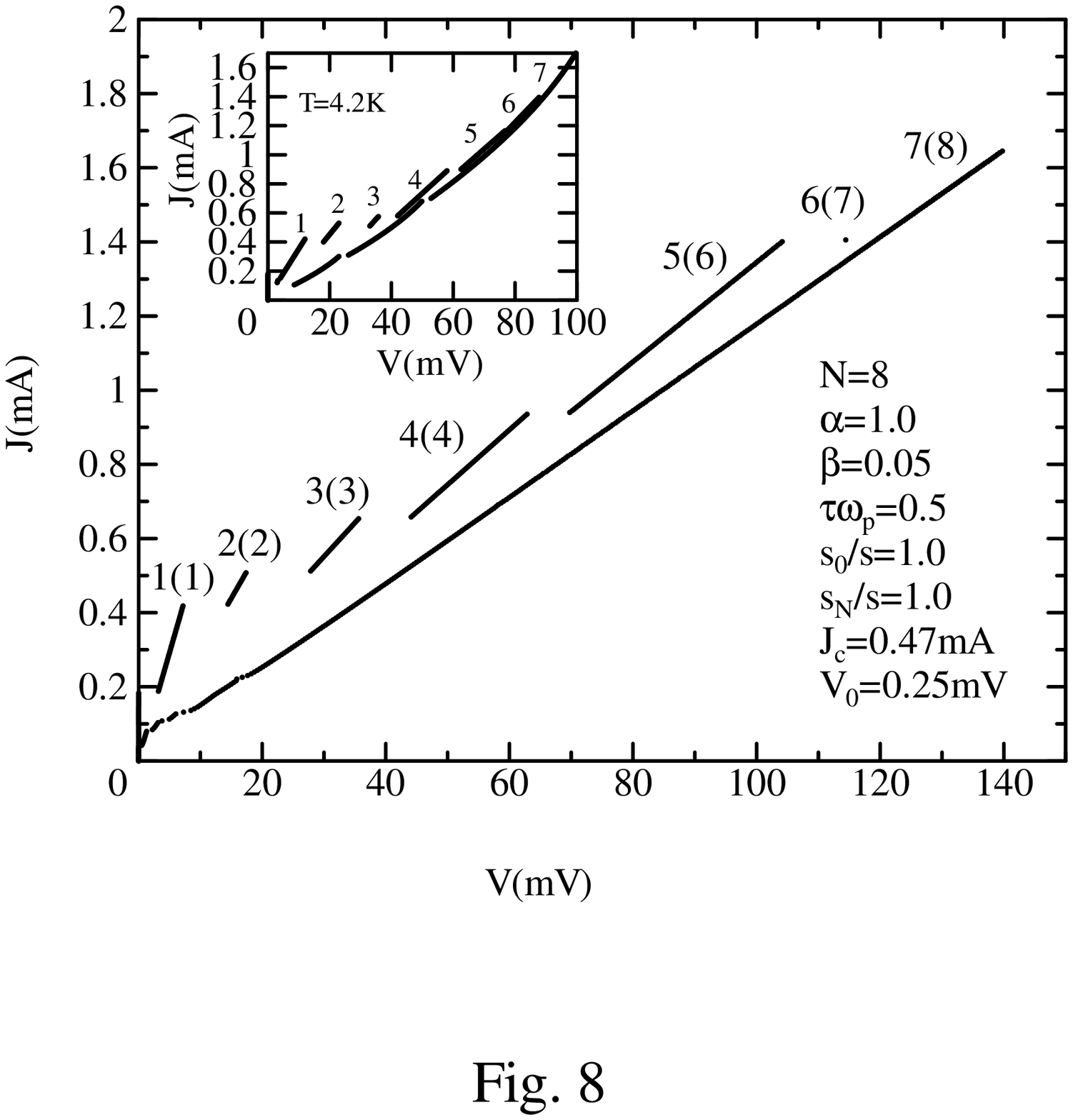}
\end{figure}

\end{document}